\documentclass[12pt]{article}
\usepackage{epsfig}
\usepackage{amssymb}

\newcommand{\R}{\mathbb{R}}
\newcommand{\C}{\mathbb{C}}

\newcommand{\fa}{\mathfrak{a}}
\newcommand{\fb}{\mathfrak{b}}
\newcommand{\fc}{\mathfrak{c}}

\newcommand{\cC}{\mathcal{C}}

\newcommand{\cH}{\mathcal{H}}

\newcommand{\cL}{\mathcal{L}}
\newcommand{\cP}{\mathcal{P}}
\newcommand{\cQ}{\mathcal{Q}}

\newcommand{\be}{\begin{equation}}
\newcommand{\ee}{\end{equation}}
\newcommand{\bea}{\begin{eqnarray}}
\newcommand{\eea}{\end{eqnarray}}

\newcommand{\kt}{\rangle}
\newcommand{\br}{\langle}

\newcommand{\ed}{\end{document}}

\newcommand{\pbr}{\prec\!}
\newcommand{\pkt}{\!\succ}

\newcommand{\bi}{\begin{itemize}}
\newcommand{\ei}{\end{itemize}}

\begin{document}

\title{Time-dependent quasi-Hermitian
Hamiltonians and the unitary quantum evolution }
\author{\\
Miloslav Znojil
\\
\\
\'{U}stav jadern\'e fyziky AV \v{C}R, 250 68 \v{R}e\v{z}, Czech
Republic\footnote{e-mail: znojil@ujf.cas.cz}}
\date{ }
\maketitle

\begin{abstract}

We show that the consequences of an introduction of a manifest
time-dependence in a pseudo-Hermitian Hamiltonian $H=H(t)$ are by
far less drastic than suggested by A. Mostafazadeh in Phys. Lett.
B 650 (2007) 208 (arXiv:0706.1872v2 [quant-ph]). In particular,
the unitarity of the evolution does not necessitate the
time-independence of the metric $\eta_+=\eta_+(t)$.

\vspace{5mm}

\vspace{2mm}


\end{abstract}

%\tableofcontents
%\textheight = 22cm \topskip = -1cm \topmargin = -1cm

\newpage

\section{Introduction}

In his letter \cite{I}, Ali Mostafazadeh arrives at a very
surprising assertion that a given time-dependent pseudo-Hermitian
Hamiltonian operator $H(t)$ defines a consistent and unitary
quantum evolution {\em if and only if} it is
\emph{quasi-stationary}, i.e., if and only if it is
$\eta_+$-pseudo-Hermitian with respect to a {\em time-independent}
metric operator $\eta_+$. In our present critical comment on this
influential letter (used, by his author, i.a., in an extremely
interesting recent discussion on the paradox of quantum
brachistochrone \cite{brasi}) we shall re-analyze the text and
show that it relies on certain assumptions which need not be
satisfied in general. In this sense we shall oppose Mostafazadeh's
conclusions and claim that the time evolution of many quantum
systems {\em can} remain unitary {\em even if} their
time-dependent pseudo-Hermitian Hamiltonian operators $H(t)$ are
left \emph{non-quasi-stationary}.

In an introductory part of our argument (section \ref{intro}) we
briefly review the terminology and summarize some basic concepts
and definitions. In the subsequent section \ref{teze} we address
``the heart of the matter" and show why the quasistationarity of
$H(t)$ as postulated in \cite{I} (and having even some practical
relevance, say, in laser physics \cite{laser}) is {\em not} a
necessary condition of the unitarity of the evolution. On an
elementary two-by-two matrix example we also demonstrate that the
assumption of the quasistationarity (i.e., of the
time-independence of the metric) is extremely counterintuitive.
Section \ref{suma} finally summarizes briefly the message of our
present comment.

\section{Quasi-Hermitian Hamiltonians \label{intro} }

%\subsection{A maze of terminology}

Scholtz et al \cite{Geyer} were probably the first {physicists}
(in fact, nuclear physicists) who discovered that whenever the
standard quantization recipe\footnote{based, say, on the principle
of correspondence} happens to produce a prohibitively complicated
version of a realistic Hamiltonian operator $h=h^\dagger$ in a
usual Hilbert space\footnote{$\cH^{(ref)}=\cL_2(\R^n)$ or a
similar ``reference" \cite{I} Hilbert space endowed with the
standard inner product $\br\cdot|\cdot\kt$}, it is still possible
to try to simplify the underlying Schr\"{o}dinger equation by its
mapping into another space. Thus, typically, a complicated
fermionic $h$ has been studied as isospectral to its simpler
bosonic partner $H$ while one maps $\cH^{(ref)}\to \cH_{phys}$ via
a mere redefinition of the inner product,
 $\br\cdot|\cdot\kt \to \br\cdot|{\bf T}\cdot\kt $.
In this context, Eq. Nr. (2) of  ref.~\cite{I} giving
 $\br\cdot|\cdot\kt \to
 \br\cdot|\cdot\kt_+:=\br\cdot|\eta_+\cdot\kt $
just updates the notation of ref.~\cite{Geyer}. An alternative
update of this type is being used by Bender et al \cite{BBJ} who
factorize the metric $\eta_+ = \cC \cP$ (where $\cP$ is parity)
and make it unique via an artificial (or, if you wish,
``physical") constraint $\cC^2=I$ imposed upon their ``charge"
operator $\cC$. In our recent comment \cite{weak} inspired by
Solombrino \cite{Solombrino} and admitting generalized
 $\cP \neq \cP^\dagger$ we showed that in such a generalization
one has either to re-write $\eta_+ = \cC \cP^\dagger$ or,
alternatively, to introduce quasiparity $\cQ$ and factorize
$\eta_+ = \cP \cQ$. Being exposed to this long menu of
alternatives we often recommend the abbreviation $\eta_+:= \Theta$
which replaces the original symbol ${\bf T}$ (too much reminiscent
of the time reversal operator of ref.~\cite{BB}) simply by its
``Greek-alphabetic" version.

In this notation, our attention  will solely be  paid here to the
quasi-Hermitian Hamiltonians \cite{Geyer} which obey the rule
given also in ref.~\cite{I} as Eq. Nr. (3),
 \be
 H^\dagger = \Theta\,H\,\Theta^{-1}\,.
 \ee
From the same source we shall also recall the subsequent Eq. Nr.
(4) re-written here, in a compactified notation with with $\omega
=\Theta^{1/2}=\eta_+^{1/2}=\omega^\dagger$, as
 \be
 h = \omega\,H\,\omega^{-1}\,.
 \label{trafo}
 \ee
This is a similarity transformation between the auxiliary
Hermitian $h=h^\dagger$ (acting in ${\cal H}^{(aux)}$) and the
quasi-Hermitian physical $H \neq H^\dagger$ (acting in ${\cal
H}_{phys}$). This mapping is unitary (cf. footnote Nr. 5 in
\cite{I}).

\subsection{A  two-by-two matrix example}

In Eqs. Nr. (17) and (18) of ref. \cite{I} a complex two-by-two
matrix Hamiltonian has been chosen for illustrative purposes. Once
we omit a trivial overall shift $q\in\R$ of its spectrum $E_1=q+E$
and $E_2=q-E$ we have
    \be
        H_0=\left(\begin{array}{cc}\fa & \fb\\ \fc & -\fa
    \end{array}\right)
 =E\left(\begin{array}{cc}\cos\theta & e^{-i\varphi}
    \sin\theta\\ e^{i\varphi}
    \sin\theta & -\cos\theta\end{array}\right),
    \label{H=}\,
    \ee
with the real scale factor $E:=\sqrt{\fa^2+\fb\fc}\in [0,\infty)$
and with the two complex angles $\theta,\varphi\in\C$ (as in
\cite{I} one could set $\Re(\theta)\in[0,\pi]$ and
$\Re(\varphi)\in[0,2\pi)$ where symbols $\Re(\cdot)$ and
$\Im(\cdot)$ denote the real- and imaginary-part functions,
respectively). Closed formulae for eigenvectors are also available
and define the positive-definite metric (cf. Eqs. Nr. (19), (20)
and (21) of \cite{I}).

For our present purposes it will be sufficient to consider just a
real-matrix subfamily of eq.~(\ref{H=}) with
$q=\Re(\theta)=\Im(\varphi)=0$ and with $\varphi=\pi/2$. This
establishes the correspondence between formulae of \cite{I} and
their special cases in \cite{Hendrik}. In the resulting reduced,
one-parametric family of Hamiltonians we have the purely imaginary
$\theta$s so that we may set $E=1/\cos \theta=\sin \alpha$ and
arrive at the toy Hamiltonian defined in terms of a single  real
variable $\alpha \in (0,\pi/2)$,
    \be
    H_{00}=\left(\begin{array}{cc}1&
    \cos\alpha\\- \cos\alpha & -1\end{array}\right)\,
    \label{H0-zero}
    \ee
(cf. \cite{Hendrik} for more details).  The eligible metrics
remain two-parametric and have a compact form
    \be
    \Theta=\Theta(H_{00})=Z\,
    \left(\begin{array}{cc}1+\sin\alpha\,\sin\gamma&
    -\cos\alpha\\- \cos\alpha & 1-\sin\alpha\,\sin\gamma
    \end{array}\right)\,.
    \label{uH0-zero}
    \ee
Although the scale factor $Z\in\R$ itself can be understood as
less relevant in the time-independent case \cite{erratum}, it is
necessary to pick up and fix a suitable value of the real angle
$\gamma \in [0,\pi/2)$. Its ambiguity is an unpleasant problem
\cite{Geyerbe}. Fortunately, the solution is easy for the
finite-dimensional Hamiltonians $H$ where one simply requires the
validity of the quasi-Hermiticity condition for some {\em other}
operators of observables ${\cal O}={\cal O}_n$ \cite{Geyer},
 \be
 {\cal O}_n^\dagger = \Theta\,{\cal O}_n\,\Theta^{-1}\,,
 \ \ \ \ \ \ \ \ n =
1,2,\ldots,N\,.
 \label{oshidashi}
 \ee
As long as one has $N=1$ in our two-dimensional real model
(\ref{H0-zero}), we may set
 \be
 {\cal O}=
 \left (
 \begin{array}{cc}
 a&b\\c&d
 \end{array}
 \right )
 \label{auxilobv}
 \ee
(with, for simplicity, real elements) and convert
eq.~(\ref{oshidashi}) in the single constraint
 \be
 (d-a)\,\cos \alpha = (b-c) + (b+c)\,\sin \alpha \sin \gamma\,.
 \label{kachikoshi}
 \ee
This enables us to fix $\gamma = \gamma(a,b,c,d)$ whenever we
choose $b\neq -c$ in our auxiliary non-Hamiltonian observable
${\cal O}$. In particular, for illustration purposes we may select
 \be
 {\cal O}_{00}=
 \left (
 \begin{array}{cc}
 0&b\\c&0
 \end{array}
 \right )\,,\ \ \ \ \ \ \
 b = \frac{1}{2}\,e^{-\varrho}\,,\ \ \ \ \ \ \
 c = \frac{1}{2}\,e^\varrho\,, \ \ \ \ \ \varrho \in (0, \infty)
  \label{auxilo}
 \ee
with real eigenvalues $=\pm 1$ (cf. also \cite{cc} in this
respect) and with eqs.~(\ref{oshidashi}) or (\ref{kachikoshi})
reduced to the single elementary relation
 \be
 \tanh \varrho = \sin \alpha \sin \gamma\,
 \label{make}
 \ee
with an easy re-insertion in eq.~(\ref{uH0-zero}).

\subsection{Bases in any number of dimensions }

%\subsection{A suppression of some hidden ambiguities in the notation }

The change of name $\eta_+ \to \Theta$  emphasizes that we
restrict attention to the quasi-Hermitian models where there
exists the positive-definite metric $\Theta$\footnote{i.e.,
equivalently, where the energies are real and {\em observable}
\cite{I}}. In comparison, the Mostafazadeh's selection of a
broader class of pseudo-Hermitian $H$s in \cite{I} seems formal,
especially when just the unitarity of the quantum evolution is
concerned and studied\footnote{as emphasized in \cite{I}, {\em
all} the quantum models become unphysical for {\em all} the
non-quasi-Hermititan pseudo-Hermitian $H$s. Beyond the domain of
quantum theory, these operators still find applications, {\it pars
pro toto}, in classical magnetohydrodynamics \cite{Uwe} }. In a
way emphasized in the context of laser optics \cite{laser} we
may/shall simply ignore the existence of a pseudometric (denoted
by symbol $\eta$ entering Eq. Nr. (1) in \cite{I}) and assume the
reality of the spectrum.

For a certain enhancement of clarity we shall employ here an
amended Dirac's notation \cite{notation} and denote the
Mostafazadeh's specific, mutually biorthogonal eigenvectors
$|\,\psi_n\kt$ and $|\,\phi_n\kt$ as the single-ketted $|\,E_n\kt$
and double-ketted $|\,E_n\kt\!\kt$, respectively. This enables us
to omit the redundant Greek letters and to rewrite Eqs. Nr. (5) of
\cite{I} in the more transparent form
 \be
 H\,|\,E_n\kt=E_n\,|\,E_n\kt\,,\ \ \ \ \ \ \ \
 H^\dagger\,|\,E_n\kt\!\kt=E_n\,|\,E_n\kt\!\kt\,.
 \label{5}
 \ee
This emphasizes the difference between $H$ and $H^\dagger$ and the
asymmetry between the two symbols since $|\,E_n\kt\!\kt \sim
\Theta \,|\,E_n\kt$ where, by assumption, $\Theta \neq I$ is
nontrivial \cite{Geyer}.

One immediately concludes that the symbol $\pbr a,b\pkt$ of
ref.~\cite{I} should in fact be re-read, in our present notation,
as the overlap $\br\!\br a\,|\,b\kt$ of the two states in the
self-dual ${\cal H}_{phys}$ where the self-duality is nontrivial,
$\Theta \neq I$. In the language of physics one could treat the
kets $|\,E_n\kt$ simply as ``elements" of ${\cal H}_{phys}$ while
the ketkets $|\,E_n\kt\!\kt$ should be understood as linear
functionals in the same Hilbert space.

In the biorthogonal-basis representation of this space  ${\cal
H}_{phys}={\cal H}^{(\Theta)}$ with $\Theta \neq I$, our present
modification of the notation enables us to make the respective
biorthogonality and completeness relations (cf. Eqs. Nr. (6) and
(7) of \cite{I}) more explicit and much more transparent,
 \be
 \br\!\br E_m\,|\,E_n\kt =\delta_{mn}\,,
 \ \ \ \ \ \ \ \ \ \
 \sum_{n=0}^\infty\,|\,E_n\kt\,\br\!\br E_n\,|
  =I\,.
 \ee
One of the important merits of this notation convention is that it
emphasizes that there exists a sequence of numbers $\kappa_n \in
\C\setminus\{0\}$ which are arbitrary free parameters. Their
existence is characteristic for biorthogonal bases as it reflects
the freedom\footnote{not existing in orthogonal bases} of a change
of the normalization of their individual elements (cf. Eqs. Nr.
(19) and (20) in \cite{I} or related remarks in our recent
preprint \cite{role}). Precisely these parameters also enter the
general formula
 \be
 \Theta=\Theta^{(\vec{\kappa})}
 =\sum_{n=0}^\infty\,|\,E_n
 \kt\!\kt
 \,\frac{1}{\kappa_n^*\kappa_n}\,
 \br\!\br E_n\,|
 \,
 \label{metrikas}
 \ee
which assigns a menu of eligible metric operators to a given $H$.
In this sense, one simply assumes that in Eq. Nr. (7) of \cite{I}
we fix the choice of the basis {\em and} of the metric {\em at
once}. Only then we are allowed to set all $\kappa_n=1$ in
(\ref{metrikas}).

\section{Evolution in time  \label{teze} }

\subsection{The amendment of the evolution law }

Let us start from a remark that, originally and paradoxically, the
evolution law based on Eqs. Nr. (10) of ref.~\cite{I} attracted
our attention not so much by its central position and by its key
relevance for the flow of the argumentation in ref.~\cite{I} but
rather by a certain innocent-looking apparent inconsistency of the
notation where in the elements $|\,\psi(t)\kt$ and $|\,\phi(t)\kt$
of a Hilbert space the kets were omitted. Suddenly (e.g., in the
footnote Nr. 5 in ref.~\cite{I}), the ``evolving state vectors"
were written and presented as the mere unbracketed functions
$\psi(t)$ and $\phi(t)$, respectively.

Our almost exaggerated attention paid to the notation helped us to
reveal that the interpretation of these kets is in fact deeply
ambiguous. Our point can immediately be clarified in a better
notation where one recollects that in \cite{I}, {\em both}
$\psi(t)$ and $\phi(t)$ were presented as elements of ${\cal
H}_{phys}$ (cf. footnote Nr. 5 in \cite{I} once more). This is in
conflict with the fact that the formalism is already presented
using a biorthogonal basis (cf. Eqs. Nr. (5), (6) and (7) in
\cite{I} or our remark made in the sequel of eq.~(\ref{5}) above).
One concludes that although just one state is prepared at an
initial time $t=0$, it can be represented in the two different
forms of a superposition over our basis. For the purposes of our
forthcoming analysis  this suggests the following change of the
denotation of the symbols entering Eq. Nr. (10) in \cite{I},
 \[
 \psi(0) = \sum_{n=0}^\infty\,|\,E_n(0)\kt\,c_n(0):= |\,\psi(0)\kt\,,
 \  \ \ \ \ \ \ \ \ \
 \phi(0)  =
 \sum_{n=0}^\infty\,|\,E_n(0)\kt\!\kt\,d_n(0):= |\,\phi(0)\kt\!\kt\,.
 \]
Moreover, once we abolished the Mostafazadeh's ``obligatory"
assignment of letters (with his $\psi_n$ meaning our states
$|\,E_n\kt$ and with his $\phi_n$ meaning the corresponding linear
functionals $|\,E_n\kt\!\kt$), we may also replace his
time-dependent symbol $\pbr \psi(t),\phi(t) \pkt$ by its present
equivalent $\br\!\br \psi(t)\,|\,\psi(t) \kt$. Its form properly
emphasizes that we consider {\em the single} time-dependent and
evolving physical state $\psi(t)$ possessing the two
mathematically slightly different ``left and right" or ``brabra
and ket" or ``functional and vector" representants
$|\,\psi(t)\kt\!\kt$ and $|\,\psi(t)\kt$, respectively.

On this background we may turn attention to $t>0$ and consider the
symbol $\pbr a,b\pkt$ of ref.~\cite{I} with $a$ replaced by the
evolving ``left ketket" and with $b$ representing the evolving
``right ket". In such an arrangement it is obvious that for $H
\neq H(t)$, the time evolution of both of these {\em different}
representations of the {\em same} state (denoted, in order to
avoid confusion, by the new symbol $\Phi$) must be controlled by
the {\em different} operators, viz., by $H$ and $H^\dagger$,
respectively. This observation simply {\em discourages} us to
postulate the evolution law in the oversimplified form of Eqs. Nr.
(10) of \cite{I}. These equations {\em must} be replaced by a more
flexible {\em double} ansatz of ref. \cite{which}, with the
different time-evolution operators acting to the right and to the
left, respectively,
 \be
 |\Phi(t)\kt=U_R(t)\, |\Phi(0)\kt\,,
 \ \ \ \ \ \ \ \ \ \
 \br\!\br \Phi(t)\,|=  \br\!\br \Phi(0)\,|\,U_L(t)\,.
   \ee
In the next step let us recollect, once more, the pullback
relations mentioned in the footnote Nr.~5 of \cite{I} and having
the compact form of a definition (\ref{trafo}) of an isospectral
Hamiltonian $h=h^\dagger$ acting in ${\cal H}^{(aux)}$. Under a
{\em specific} normalization, our biorthogonal basis in
$\cH_{phys}={\cal H}^{(\Theta)}$ may be assumed adapted, as we
already agreed, to a fixed $\Theta$ in such a way that
$\kappa_n=1$ at all $n$ in eq.~(\ref{metrikas}) (cf. Eq. Nr. (7)
in \cite{I}). In this way we re-derive the spectral-representation
formula
 \be
 h=\sum_{n=0}^\infty\,|\chi_n\kt\,E_n\,\br \chi_n\,|=h^\dagger\,,
 \ \ \ \ \ \ \ |\,\chi_n\kt = \omega\,|\,E_n\kt =
 \omega^{-1}\,|\,E_n\kt\!\kt\,.
 \ee
The textbook wisdom becomes applicable and we can immediately
deduce the time-evolution law in ${\cal H}^{(aux)}$. Thus, at any
time-dependence in $h=h(t)$ the evolution starts from the state
 \[
 |\,\chi(0)\kt =
 \sum_{n=0}^\infty\,\omega(0)\,|\,E_n(0)\kt\,c_n(0)= \omega(0)
 \,|\,\psi(0)\kt\,
 \equiv\,
  \sum_{n=0}^\infty\,\omega^{-1}(0)\,|\,E_n(0)\kt\!\kt\,d_n(0)
  = \omega^{-1}(0)\,|\,\phi(0)\kt\!\kt\,
 \]
prepared in ${\cal H}^{(aux)}$ at $t=0$. There is no problem with
writing down its descendant existing at $t>0$,
 \[
 |\,\chi(t)\kt = u(t)\,|\,\chi(0)\kt\,.
 \]
Here, the evolution operator is determined by the standard
Schr\"{o}dinger equation,
 \be
 {\rm i}\partial_t u(t)=h(t)\,u(t)\,,\ \ \ \ \ \ \
 u(0)=I\,.
 \label{seh}
 \ee
It is easy to deduce that
 \[
 |\,\Phi(t)\kt = \omega^{-1}(t)\, |\,\chi(t)\kt =
 U_R(t)\,|\,\Phi(0)\kt\,,
 \]
 \[
 |\,\Phi(t)\kt\!\kt = \omega(t)\, |\,\chi(t)\kt =
 U_L^\dagger(t)\,|\,\Phi(0)\kt\!\kt\,.
 \]
With the insertions of
$|\,\Phi(0)\kt=\omega^{-1}(0)\,|\,\chi(0)\kt$ and
$|\,\Phi(0)\kt\!\kt=\omega(0)\,|\,\chi(0)\kt$ we may conclude that
 \be
 U_R(t)=\omega^{-1}(t)\,u(t)\,\omega(0)\,,\ \ \ \ \ \ \
 U_L^\dagger(t)=\omega^\dagger(t)\,u(t)\,
 \left [\omega^{-1}(0)\right ]^\dagger\,.
 \label{evolop}
 \ee
This means, obviously, that
 \be
 \br\!\br \Phi(t)\,|\,\Phi(t) \kt=
 \br\!\br \Phi(0)\,|\,U_L(t)\,U_R(t)
 \,|\Phi(0) \kt=
 \br\!\br \Phi(0)\,|\,\Phi(0) \kt\,.
 \ee
We demonstrated that the evolution is unitary.

\subsection{Two-by-two model made time-dependent}

In a two-level model let us introduce time $t$ {\em not only} in
the single-parametric Hamiltonian,  by admitting that
$\alpha=\alpha(t)$ in $H_{00}$ of eq.~(\ref{H0-zero}), {\em but
also } in the complementary observable ${\cal O}_{00}$,  by
allowing that $\varrho=\varrho(t)$ in eq.~(\ref{auxilo}). This
gives us the basic methodical guidance for a replacement of
eq.~(\ref{trafo}) by its time-dependent generalization
 \be
 h(t) = \omega(t)\,H(t)\,\omega^{-1}(t)\,.
 \label{traform}
 \ee
In full analogy with the time-independent case \cite{Geyer} we
imagine that the operators on both sides of eq.~(\ref{traform})
should represent the same information about the dynamics of our
system. This means that we are allowed to ignore the specific
additional quasi-stationarity constraints deduced from the
incorrect assumptions in \cite{I}. One comes to this conclusion
with a great relief since in our example the time-dependence
encoded in $\alpha(t)$ and $\varrho(t)$ becomes immediately
transferred, via eq.~(\ref{make}), to the metric $\Theta$ in
eqs.~(\ref{uH0-zero}) and to its square root $\omega$ in
eq.~(\ref{auxhh}).

One of the specific merits of our choice of the example is that
the latter two matrices can still be written in closed form.
Indeed, in terms of the independently variable  $T=\tanh
\varrho=T(t)$ and $C=\cos \alpha=C(t)$ we may define our metric
$\Theta=\Theta(t)$ as well as the matrix $U=U(t)$ of its
(unnormalized) eigenvectors by the elementary prescriptions
 \[
 \Theta=Z\,
 \left (
 \begin{array}{cc}
 1+T&-C\\-C&1-T
 \end{array}
 \right )\,,\ \ \ \ \ \ \ \ \ \
 U=
 \left (
 \begin{array}{cc}
 T+R&C\\-C&T+R
 \end{array}
 \right )\,.
 \]
Using further the abbreviations $R=\sqrt{T^2+C^2}$ and
 \[
 S=\,\frac{2R}{C^2+(T+R)^2}\,\times\,\frac{1}{\sqrt{1-R}+\sqrt{1+R}}\,
 \]
it is entirely straightforward to derive
 \be
 \omega=
 \left (
 \begin{array}{cc}
 \sqrt{1-R}+SC^2&-SC(T+R)\\-SC(T+R)&\sqrt{1-R}+S(T^2+R^2)
 \end{array}
 \right )\,
 \label{auxhh}
 \ee
i.e., the closed form of the matrix of transformation entering
eqs.~(\ref{trafo}) and (\ref{evolop}).

\section{Conclusions \label{suma} }

It is obvious that once {\em both} a quasi-Hermitian Hamiltonian
$H$ {\em and} the associated observables ${\cal O}_n$ become {\em
manifestly and simultaneously} time-dependent, one encounters an
entirely new situation because the time-dependence of the system
becomes, in general, transferred to the metric $\Theta$. This
means that in principle, the evolution of the quantum system
ceases to be dictated solely by the Hamiltonian.

In an opposite direction, once one makes a {\em tacit} assumption
that the metric $\Theta(t)$ {\em does not} carry any independent
information about the changes of dynamics with time, no trace of
the  variation of the associated observables ${\cal O}_n={\cal
O}_n(t)$ would be left observable. We may conclude that this type
of assumption does not seem reasonable at all. Moreover,
ref.~\cite{I} remains useful as showing how such an assumption
would produce severe restrictions imposed upon $H(t)$ itself.

We re-analyzed the problem since we were really unpleasantly
surprised by the drastic nature of the conditions of unitarity of
the time evolution as deduced in \cite{I}. Using the
modified\footnote{viz, brabraket \cite{notation}} Dirac's notation
we re-derived and corrected the key Eqs. Nr. (10) of
ref.~\cite{I}.

Fortunately, the situation is clarified now. Our conclusion is
that even for the quasi-Hermitian models there arise no problems
with the unitarity of the evolution. Briefly it is possible to
summarize that whenever we have $H=H(t)$ {\em and}
 ${\cal O}_n={\cal O}_{n}(t)$, we have to search for an evolution
equation which depends {\em not only} on the Hamiltonian $H(t)$
but also on the changes in time which are carried by the metric
$\Theta=\Theta(t)$ itself. In another perspective, in contrast to
the hypotheses formulated in ref.~\cite{I}, there emerge no
surprising and far reaching differences between the role and/or
interpretation of the time-dependent and time-independent
pseudo-Hermitian Hamiltonians.

\section*{Acknowledgement}

Work supported by GA\v{C}R, grant Nr. 202/07/1307, Institutional
Research Plan AV0Z10480505 and by the M\v{S}MT ``Doppler
Institute" project Nr. LC06002.

%==========copy again===========

\ed

%
%======================
%
%\bibitem{ea}
%M. Znojil, Math. Reviews, to appear (a concise ``extended
%abstract" of paper \cite{I}, custom-made for AMS, with text
%available also at http://gemma.ujf.cas.cz/\~ znojil/mr[4230].txt).
%
%